\documentclass[preprint,prd,superscriptaddress,amsmath,nofootinbib]{revtex4}
 
\usepackage{cancel}
\usepackage{color}
\usepackage{graphicx}
\usepackage{amssymb}

\def\sq{\ifmmode{\tilde{q}} \else{$\tilde{q}$} \fi}
\def\st{\ifmmode{\tilde{t}} \else{$\tilde{t}$} \fi}
\def\sg{\ifmmode{\tilde{g}} \else{$\tilde{g}$} \fi}
\def\tchi{\ifmmode{\tilde{\chi}} \else{$\tilde{\chi}$} \fi}
\newcommand{\bea}{\begin{eqnarray}}
\newcommand{\eea}{\end{eqnarray}}

\newcommand{\eq}[1]{eq.~(\ref{#1})}

\newcommand{\be}{\begin{equation}}
\newcommand{\ee}{\end{equation}}

\begin{document}
\preprint{\vbox{\hbox{PSI-PR-17-02}}}
\preprint{\vbox{\hbox{TU-1043}}}

\title{The MSSM without Gluinos; an Effective Field Theory for the Stop Sector \bigskip}

\author{Jason Aebischer} \email{aebischer@itp.unibe.ch}
\affiliation{Albert Einstein Center for Fundamental Physics, Institute
  for Theoretical Physics,\\ University of Bern, CH-3012 Bern,
  Switzerland.}
\author{Andreas Crivellin} \email{andreas.crivellin@cern.ch}
\affiliation{Paul Scherrer Institut, CH--5232 Villigen PSI, Switzerland}
\author{Christoph Greub}\email{greub@itp.unibe.ch}
\affiliation{Albert Einstein Center for Fundamental Physics, Institute
  for Theoretical Physics,\\ University of Bern, CH-3012 Bern,
  Switzerland.}
\author{Youichi Yamada}\email{yamada@tuhep.phys.tohoku.ac.jp}
\affiliation{ Department of Physics, Tohoku University, Sendai 980-8578, Japan}
\bigskip
\date{\today\bigskip\bigskip}

\begin{abstract}
In this article we study the MSSM with stops and Higgs scalars much lighter than gluinos and squarks of the first two generations. In this setup, one should use an effective field theory with partial supersymmetry in which the gluino and heavy squarks are integrated out in order to connect SUSY parameters (given at a high scale) to observables in the stop sector. In the construction of this effective theory, valid below the gluino mass scale, we take into account $O(\alpha_3)$ and $O(Y_{t,b}^2)$ effects and calculate the matching as well as the renormalization group evolution. As a result, the running of the parameters for the stop sector is modified with respect to the full MSSM and SUSY relations between parameters are broken. We show that for some couplings sizable numerical differences exist between the effective field theory approach and the naive calculation based on the MSSM running.
\end{abstract}

\maketitle

\section{Introduction}
\label{sec:intro}

There are several theoretical arguments for a light stop in supersymmetric theories. Foremost, in natural supersymmetry (SUSY) light stops are required to cancel the quadratic divergence of the Higgs mass originating from the self-energy involving a top quark, while the other supersymmetric partners can be much heavier~\cite{Dimopoulos:1995mi,Giudice:2004tc} due to the smaller couplings to the Higgs. Moreover, the renormalization group equations (RGE) of the minimal supersymmetric standard model (MSSM) generically drive the bilinear mass term parameters of the third generation squarks to lower values (compared to the first two generations) due to their non-negligible Yukawa couplings~\cite{Inoue:1982pi,Inoue:1983pp,Derendinger:1983bz,Gato:1984ya,Falck:1985aa,Martin:1993zk}. 
\medskip

Although the measured Higgs mass of around 125 GeV~\cite{Aad:2012tfa,Chatrchyan:2012ufa} prefers rather heavy (around the TeV scale)~\cite{Ellis:1990nz,Okada:1990vk,Haber:1990aw} rather than light stops in the MSSM, this is not necessarily the case in the NMSSM~\cite{Ellwanger:2009dp}, in $\lambda$SUSY models \cite{Hall:2011aa}, models with light sneutrinos \cite{Chala:2017jgg} or in supersymmertic models with additional D-term \cite{Batra:2003nj} or F-term \cite{Espinosa:1998re} contributions to the scalar potential. Also large (or even maximal \cite{Djouadi:2005gj,Brummer:2012ns,Wymant:2012zp}) stop mixing angles help to get the right Higgs mass with rather light stops. 
\medskip

LHC searches for top squarks (using simplified models) set a lower bound on its mass of around $m_{\tilde t_1}=300\,\text{GeV}$, which however heavily depends on the neutralino mass. Depending on the stop and the neutralino mass, different decay modes are studied. For the decay channel $\tilde t_1\rightarrow t\,\tilde \chi^0_1$ \cite{Aad:2014qaa,ATLAS-CONF-2016-050,Chatrchyan:2013xna}, the limits are quite stringent, even though for light neutralinos very light stops can not be excluded due to the high $t\bar t$-background \cite{Cheng:2016mcw}. The three-body decay $\tilde t_1\rightarrow Wb\tilde \chi^0_1$ was analyzed theoretically in \cite{Porod:1996at} and experimentally in \cite{ATLAS-CONF-2016-076}. Finally the decay $\tilde t_1\rightarrow c\,\tilde \chi^0_1$ and the less important four-body decay $\tilde t_1\rightarrow \,\tilde \chi^0_1d_if\bar f'$ are treated in \cite{Aaboud:2016tnv,Aebischer:2014lfa,Grober:2014aha} and constraints were derived by the ATLAS collaboration from the monojet
analysis in ~\cite{Aad:2014nra}. Some bounds can be avoided in kinematic boundary regions or once non-minimal flavour violation is included. However, recently efforts of closing these gaps have been made \cite{Macaluso:2015wja,Belyaev:2015gna,Kobakhidze:2015scd,Crivellin:2016rdu} and stops should in general not be lighter than 300 GeV.
Nevertheless, the mass bound for the stop is still weaker than the 
strong bounds on the squark masses of the first two generations and also on the gluino mass~\cite{ATLAS:2016lsr,Sakuma:2016nxo}. 
For sbottom quarks LHC searches suggest masses of above $800\,\text{GeV}$ \cite{ATLAS:2017qih,Sirunyan:2017kiw}.
The bounds on sparticles with EW interactions only are much less stringent~\cite{Aad:2015jqa,ATLAS:2016uwq,Arina:2016rbb,Khachatryan:2014mma,CMS:2016gvu}. For example, in the case of heavy winos the Higgsino mass parameter $\mu$ has only to be larger than $350\,\text{GeV}$ \cite{Aad:2014vma}. It can be shown however that by changing the assumptions on the composition of charginos and neutralinos, collider limits can get even further weakened~\cite{Bharucha:2013epa,Martin:2014qra,Calibbi:2014lga,Chakraborti:2015mra}. For the Higgs bosons, different fits \cite{ATLAS:2014kua,Khachatryan:2014jya,Celis:2013rcs,Chiang:2013ixa,Chen:2013rba,Craig:2013hca,Wang:2014lta,Grinstein:2013npa,Han:2017pfo} suggest an alignment limit, in which the lightest CP-even Higgs boson takes the role of the SM Higgs. 
Collider limits on non-SM Higgs bosons for large values of $\tan \beta$ suggest that CP-odd Higgs bosons should be heavier than $800\,\text{GeV}$~\cite{Aaboud:2016cre,Khachatryan:2014wca}.

\medskip
If the gluino (or the squarks of the first two generations~\cite{Giudice:2004tc,ArkaniHamed:2004yi}) is much heavier than the stops, an effective theory (EFT) with partial SUSY must be constructed in which the gluino (squarks) is integrated out~\cite{Muhlleitner:2008yw} (\cite{Carena:2008rt,Giudice:2011cg}). Such a hierarchy can for example be achieved for MSSM-like models in a Scherk-Schwarz breaking scenario~\cite{Pomarol:1998sd,Dimopoulos:2014aua,Garcia:2015sfa,Delgado:2016vib}. The construction of the effective theory for the stop sector is the goal of this article. Assuming a common large mass of order $M$ for the gluino and the squarks of the first two generations, we compute the matching condition between the full MSSM and the effective theory, including one-loop contributions which are enhanced by powers of $M$. Furthermore, since some supermultiplets are partially integrated out in the effective theory, the supersymmetric relations between gauge/Yukawa couplings, gaugino/Higgsino couplings and four-scalar couplings are broken in the effective theory by radiative corrections. Therefore, these couplings in the effective theory have an independent renormalization group evolution, as discussed in \cite{Chankowski:1989du,Hikasa:1995bw,Nojiri:1996fp,Cheng:1997sq,Cheng:1997vy,Nojiri:1997ma,Katz:1998br,Kiyoura:1998yt,Muhlleitner:2008yw} mainly for the gaugino-matter couplings.  
\medskip

This article is structured as follows: In the next section we establish our effective theory for the stop sector and calculate the matching as well as the running of the relevant parameters at order $\alpha_3=g_3^2/(4\pi)$, $Y_t^2$ and $Y_b^2$ (neglecting $O(g_1^2)$, $O(g_2^2)$ and Higgs self-coupling effects). This section is followed by a numerical analysis in Sec.~\ref{sec:numerics}. Finally we conclude in Sec.~\ref{sec:conclusions}.
\medskip

\section{The effective theory for the stop sector}

The aim of this section is to construct the effective theory for the MSSM stop sector, including $O(\alpha_3,Y_{t,b}^2)$ and enhanced effects. As noted before, we assume that the gluino and the squarks of the first two generations are much heavier, with masses of the order $M$, than the stops, the Higgs scalars and the Higgsinos. The left-handed sbottom is also assumed to be light such that it remains in the effective theory, forming an $SU(2)$ multiplet with the left-handed stop. However, we assume that the right-handed sbottom is heavy, with the mass of the order $M$. Therefore, we consider the following effective Lagrangian which is valid below the scale $M$, 
\begin{eqnarray}
{\cal L}_{\rm eff} &=& {\cal L}_K 
- \bar m_2^2 H_u^{\dagger}H_u - \bar m_1^2 H_d^{\dagger}H_d - V\left(H_u, H_d\right) 
\nonumber \\
&&
+ \bar m_{12}^2 H_d\cdot H_u- \bar \mu\tilde{H}_U\cdot\tilde{H}_D +(h.c.)
\nonumber \\ 
&& 
-\bar m_{\tilde{Q}}^2 \sq_L^{\dagger}\sq_L - \bar m_{\st}^2 \st^{\dagger}_R\st_R 
\nonumber \\  
&& -\bar{Y}_t \bar{t}_R q_{3L}\cdot H_u -\bar{Y}_b\bar{b}_R H_d \cdot q_{3L} +(h.c.)
\nonumber \\ 
&&
-\lambda_1^u(\sq_L^{\dagger}\sq_L)(H^{\dagger}_uH_u)
-\lambda_2^u(\sq_L^{\dagger}H_u)(H^{\dagger}_u\sq_L)
-\lambda_3^u(\st_R^{\dagger}\st_R)(H_u^{\dagger}H_u)
\nonumber \\ 
&&
-\lambda_1^d(\sq_L^{\dagger}\sq_L)(H_d^{\dagger}H_d)
-\lambda_2^d(\sq_L^{\dagger}H_d)(H_d^{\dagger}\sq_L)
-\lambda_3^d(\st_R^{\dagger}\st_R)(H_d^{\dagger}H_d)
\nonumber \\ 
&&
-\lambda_4(\sq_{Li}^{\dagger}\sq_{Li})(\sq_{Lj}^{\dagger}\sq_{Lj})
-\lambda_5(\sq_{Li}^{\dagger}\sq_{Lj})(\sq_{Lj}^{\dagger}\sq_{Li})
\nonumber \\
&& 
-\lambda_6(\sq_{Li}^{\dagger}\sq_{Li})(\st_{R}^{\dagger}\st_{R})
-\lambda_7(\sq_{Li}^{\dagger}\st_{R})(\st_{R}^{\dagger}\sq_{Li})
-\lambda_8(\st_{R}^{\dagger}\st_{R})(\st_{R}^{\dagger}\st_{R})
\nonumber \\
&&
-\bar{A}_t\st_R^{\dagger}\sq_{L}\cdot H_u +  \bar\mu_t \tilde t_R^{\dag} H_d^{\dag}\tilde q_L +(h.c.)
\nonumber \\
&&
-\bar{Y}_{q_{3L}}\st_R^{\dagger}q_{3L}\cdot\tilde{H}_U 
-\bar{Y}_{t_R}\bar{t}_R\sq_L\cdot \tilde{H}_U 
-\bar{Y}_{b_R}\bar{b}_R \tilde{H}_D \cdot \sq_L +(h.c.)\,,
\label{Leff}
\end{eqnarray}
with partial supersymmetry. Here ${\cal L}_K$ denotes the kinetic terms and gauge interactions, and $V(H_u,H_d)$ denotes the quartic couplings of the Higgs doublets ($H_u$, $H_d$). For the interactions involving four squarks, the $SU(3)$ color indices are contracted within the parentheses. 
Similarly, the $SU(2)$ indices in the two-squark-two-Higgs interactions are contracted within the parentheses. $i,j$ are the $SU(2)$ indices and the dot denotes the contraction of $SU(2)$ indices as $A\cdot B=A_1B_2-A_2B_1$. For simplicity, we also assume that the electroweak gauginos and sleptons are heavy. However, since we neglect $O(g_1^2)$, $O(g_2^2)$ effects in the following, relaxing this assumption would leave our RGEs unchanged.
We also ignore the non-holomprphic Higgs-quark couplings $\bar{t}_RH_d^{\dag}q_{3L}$ and $\bar{b}_RH_u^{\dag}q_{3L}$ which are induced at the loop-level~\cite{Hempfling:1993kv,Hall:1993gn,Carena:1994bv,Noth:2008tw,Hofer:2009xb,Crivellin:2010er,Crivellin:2011jt,Crivellin:2012zz}.

\subsection{Tree-level matching} 

At the matching scale $M$ the Lagrangian of~\eq{Leff} has to be compared to the one of the full MSSM (see for example~\cite{Fayet:1976cr,Fayet:1976et,Haber:1984rc,Rosiek:1995kg}) which originates from the superpotential 
\begin{eqnarray}
W &=& Y_t T^c Q\cdot H_u + Y_b B^c H_d \cdot Q + \mu H_u\cdot H_d\,, 
\end{eqnarray}
the soft SUSY breaking terms
\begin{eqnarray}
V_{\rm soft} &=& m_{\tilde{Q}}^2 \sq^{\dagger}_L\sq_L 
+ m_{\st}^2 \st^{\dagger}_R \st_R+ m_{H_d}^2 H_d^{\dagger}H_d 
+ m_{H_u}^2 H_u^{\dagger}H_u +m_{\tilde b_R}^2\tilde b^{\dag}_R\tilde b_R
\nonumber\\ 
&& 
+A_t \st^{\dagger}_R \sq_{L}\cdot H_u+A_b \tilde b_R^{\dag}H_d\cdot \sq_{L}  - m_{H_d H_u}^2 H_d\cdot H_u + (h.c.)\,,
\end{eqnarray}
and the $D$ terms
\begin{equation}
	{V_D} = \frac{{{g_3^2}}}{2}{\left( {\tilde q_L^\dag {T^A}{{\tilde q}_L} - \tilde t_R^\dag {T^A}{{\tilde t}_R} - \tilde b_R^\dag {T^A}{{\tilde b}_R}} \right)^2},
\end{equation}
where $T^A$ are the generators of $SU(3)$ in the fundamental representation.

The matching conditions for the bilinear terms and the trilinear couplings are 
\begin{eqnarray}
\bar{Y}_t &=& Y_t\,, \qquad
\bar{Y}_b = Y_b\,, \qquad
\bar{Y}_{q_{3L}} = Y_t\,, \qquad 
\bar{Y}_{t_R} = Y_t\,, \qquad
\bar{Y}_{b_R} = Y_b\,, \qquad
\bar{A}_t = A_t\,,\\
 \bar{\mu} &=& \mu\,,\qquad
\bar{\mu}_{t} = \mu Y_t\,,\qquad
 \bar m_2^2 = m_{H_u}^2+\mu^2\,,\qquad
 \bar m_1^2 = m_{H_d}^2+\mu^2\,,\\
 \bar m_{12}^2 &=& m_{H_dH_u}^2\,,\qquad
\bar m_{\tilde Q}^2 =m_{\tilde Q}^2\,,\qquad
\bar m_{\tilde t}^2 =m_{\tilde t}^2\,.
\end{eqnarray}
The couplings between squarks and Higgs bosons are generated by F- and D-terms in the MSSM Lagrangian. At the scale $M$, they are given 
by 
\begin{eqnarray}
\lambda_1^u &=& Y_t^2\,,\qquad
\lambda_2^u = -Y_t^2\,,\qquad
\lambda_3^u = Y_t^2\,,\\
\lambda_1^d &=& Y_b^2\,, \qquad
\lambda_2^d = -Y_b^2\,,\qquad
\lambda_3^d = 0\,, \\
\lambda_4 &=& -\frac{1}{12}g_3^2 \,, \qquad
\lambda_5 = \frac{1}{4}g_3^2 \,,\qquad
\lambda_6 = \frac{1}{6} g_3^2 \,, \\
\lambda_7 &=& -\frac{1}{2} g_3^2 +Y_t^2 \,, \qquad
\lambda_8 = \frac{1}{6} g_3^2 \,,
\end{eqnarray}
keeping only Yukawa couplings and $g_3$.

\subsection{1-loop matching} 

For the matching, we need to include the one-loop effects enhanced by powers of $M$ since their contributions may be comparable to the tree level ones shown in the previous subsection. They can only appear in bilinear and trilinear terms, as seen by dimensional analysis.  
The bilinear terms receive the following shifts at the matching scale ${\mu_R}=M$
\begin{eqnarray}
 \Delta\bar m_2^2 &=& 0\,,\qquad
 \Delta\bar m_1^2 = -\frac{3}{16\pi^2}(Y_b^2m_{\tilde b_R}^2+A_b^2)\left(1-\log{\left(\frac{m_{\tilde b_R}^2}{M^2}\right)}\right)\,,\\
  \Delta\bar m_{\tilde Q}^2 &=& -\frac{1}{16\pi^2}(Y_b^2m_{\tilde b_R}^2+A_b^2)(1-\log{\left(\frac{m_{\tilde b_R}^2}{M^2}\right)})+\frac{\alpha_3 C_F}{\pi}m^2_{\tilde g}\left(1-\log{\left(\frac{m_{\tilde g}^2}{M^2}\right)}\right),
\end{eqnarray}
\begin{eqnarray}	
  \Delta\bar m_{\tilde t}^2 &=&\frac{\alpha_3 C_F}{\pi}m_{\tilde g}^2\left(1-\log{\left(\frac{m_{\tilde g}^2}{M^2}\right)}\right)\,,\\
 \Delta\bar m_{12}^2 &=&-\frac{3A_b\mu Y_b}{16\pi^2}\left(1-\log{\left(\frac{m_{\tilde b_R}^2}{M^2}\right)}\right)\,,\\
\Delta\bar{\mu}&=&0\,.
\end{eqnarray}
For the trilinear term the shift reads
\begin{eqnarray}
\Delta\bar{A}_t &=& -\frac{A_b Y_t Y_b}{16\pi^2}\left(1-\log{\left(\frac{m_{\tilde b_R}^2}{M^2}\right)}\right)-\frac{\alpha_3 C_F}{\pi}m_{\tilde g}Y_t\left(1-\log{\left(\frac{m_{\tilde g}^2}{M^2}\right)}\right)\,,\\
\Delta\bar{\mu}_t&=&0\,.
\label{eq:At thresh}
\end{eqnarray}
All the other parameters relevant for the stop sector are dimensionless and therefore do not receive any $M$ enhanced corrections. 

\subsection{Renormalization group evolution}

The running of the full MSSM parameters~\cite{Inoue:1982pi,Inoue:1983pp,Derendinger:1983bz,Gato:1984ya,Falck:1985aa} is known at the two-loop level~\cite{West:1984dg,Jones:1984cx,Martin:1993yx,Yamada:1993ga,Martin:1993zk}. Here we give the one-loop beta functions to $\mathcal{O}(\alpha_3,Y_{t,b}^2)$ for the parameters of our effective theory in \eq{Leff}. The corresponding results for the full MSSM are summarized in the appendix. For the strong coupling constant we have 
($t\equiv\log {\mu_R}$, where {$\mu_R$} denotes the renormalization scale) 
\begin{eqnarray}
16\pi^2\frac{d}{dt}\bar g_3 &=& \left(-7+\frac{1}{2}\right)\bar g_3^3\,,
\end{eqnarray}
where the first term on the right hand side is the SM contribution. The effective quark-quark-Higgs Yukawa couplings evolve according to
\begin{eqnarray}
16\pi^2\frac{d}{dt}\bar{Y}_t &=& \bar{Y}_t \left[ 
-8\bar g_3^2 +\frac{9}{2}\bar{Y}_t^2 +\frac{1}{2} \bar{Y}_b^2  +\bar{Y}_{t_R}^2 + \frac{1}{2} \bar{Y}_{q_{3L}}^2 \right] \,, 
\\
16\pi^2\frac{d}{dt}\bar{Y}_b &=& \bar{Y}_b \left[-8\bar g_3^2 +\frac{1}{2}\bar{Y}_t^2 + \frac{9}{2}\bar{Y}_b^2  + \bar{Y}_{b_R}^2 + \frac{1}{2} \bar{Y}_{q_{3L}}^2 \right] \,,
\end{eqnarray}
while the evolution of the ones entering the Higgsino-quark-squark vertex is determined by
\begin{eqnarray}
16\pi^2\frac{d}{dt} \bar{Y}_{q_{3L}} &=& 
\bar{Y}_{q_{3L}} \left[ -4\bar g_3^2 +\frac{1}{2}\bar{Y}_t^2+\frac{1}{2}\bar{Y}_b^2 +4\bar{Y}_{q_{3L}}^2 + \frac{3}{2} \bar{Y}_{t_R}^2 \right] \,, 
\\
16\pi^2\frac{d}{dt} \bar{Y}_{t_R} &=& 
\bar{Y}_{t_R} \left[ -4\bar g_3^2 +\bar{Y}_t^2 +\frac{3}{2}\bar{Y}_{q_{3L}}^2 + \frac{7}{2} \bar{Y}_{t_R}^2 +\bar{Y}_{b_R}^2 \right] \,, 
\\
16\pi^2\frac{d}{dt} \bar{Y}_{b_R} &=& \bar{Y}_{b_R} \left[ -4\bar g_3^2  +\bar{Y}_b^2 +\bar{Y}_{t_R}^2 
+ \frac{7}{2} \bar{Y}_{b_R}^2 \right] \,. 
\end{eqnarray}

For the Higgs mass parameters we find
\begin{eqnarray}
16\pi^2\frac{d}{dt} \bar m_2^2 &=&  
6\bar{Y}_t^2 \bar m_2^2 +6(2\lambda^u_1+\lambda^u_2)\bar m_{\tilde{Q}}^2 + 6\lambda^u_3\bar m_{\st}^2 
+ 6 \bar{A}_t^2\,,
\\
16\pi^2\frac{d}{dt} \bar m_1^2 &=& 6 \bar{Y}_b^2 \bar m_1^2+6(2\lambda^d_1+\lambda^d_2)\bar m_{\tilde{Q}}^2 + 6\lambda^d_3\bar m_{\st}^2 + 6 \bar{\mu}_t^2 \,,
 \\ 
16\pi^2\frac{d}{dt}\bar m_{12}^2 &=&  
3(\bar{Y}_t^2+  \bar{Y}_b^2 )\bar m_{12}^2 +6 \bar \mu_t \bar A_t\,,
\end{eqnarray} 
and for the bilinear squark mass terms
\begin{eqnarray}
16\pi^2 \frac{d}{dt} \bar m_{\tilde{Q}}^2 &=& 
\left[ -8\bar g_3^2 +2\bar{Y}_{t_R}^2 +2\bar{Y}_{b_R}^2 + 28\lambda_4 +20\lambda_5 \right] \bar m_{\tilde{Q}}^2 
 +(6\lambda_6+2\lambda_7)\bar m_{\st}^2 
\nonumber \\ 
&&+(4\lambda_1^u+2\lambda_2^u) \bar m_2^2 +(4\lambda^d_1+2\lambda^d_2) \bar m_1^2 
\nonumber \\ && 
+2(\bar{A}_t^2+\bar{\mu}_t^2) -4(\bar{Y}_{t_R}^2+\bar{Y}_{b_R}^2)\bar \mu^2 ,
\\
16\pi^2 \frac{d}{dt} \bar m_{\st}^2 &=& 
\left[ -8\bar g_3^2 +4\bar{Y}_{q_{3L}}^2 +16\lambda_8 \right] \bar m_{\st}^2 +(12\lambda_6+4\lambda_7)\bar m_{\tilde{Q}}^2
\nonumber \\ 
&& +4\lambda^u_3 \bar m_2 +4\lambda^d_3 \bar m_1 
+4(\bar{A}_t^2+\bar{\mu}_t^2) -8\bar{Y}_{q_{3L}}^2 \bar \mu^2 .
\end{eqnarray}

The Higgsino mass in the effective theory evolves as
\begin{equation}
16\pi^2 \frac{d}{dt} \bar \mu =  \frac{3}{2}(\bar{Y}_{q_{3L}}^2 + \bar{Y}_{t_R}^2 +\bar{Y}_{b_R}^2 )
 \bar \mu\,,
\end{equation}
and the effective trilinear $H\sq\sq$ coupling as
\begin{eqnarray}
16\pi^2\frac{d}{dt}\bar{A}_t &=& 
\bar{A}_t \left[ -8\bar g_3^2 + 2\bar{Y}_{q_{3L}}^2 +\bar{Y}_{t_R}^2 +\bar{Y}_{b_R}^2 
 +3\bar{Y}_t^2+2\lambda^u_1 -2\lambda^u_2 +2\lambda^u_3 +2\lambda_6 +6\lambda_7 \right] ,
\\
16\pi^2\frac{d}{dt}\bar{\mu}_t &=& 
\bar{\mu}_t \left[ -8\bar g_3^2 + 2\bar{Y}_{q_{3L}}^2 +\bar{Y}_{t_R}^2 +\bar{Y}_{b_R}^2 +3\bar{Y}_b^2+2\lambda^d_1 
+4\lambda^d_2 
+2\lambda^d_3 +2\lambda_6 +6\lambda_7 \right] 
\nonumber \\
&& + 4 \bar{Y}_{q_{3L}}\bar{Y}_{b_R}\bar{Y}_b \bar{\mu}.
\end{eqnarray}

Finally for the quartic $HH\sq\sq$ and $\sq\sq\sq\sq$ couplings one obtains
\begin{eqnarray}
16\pi^2\frac{d}{dt} \lambda_1^u &=& 
4(\lambda_1^u)^2 +2(\lambda_2^u)^2
+28\lambda_1^u\lambda_4 +20\lambda_1^u\lambda_5 +12\lambda_2^u\lambda_4 + 4\lambda_2^u\lambda_5 +6\lambda_3^u\lambda_6 
\nonumber \\
&& +2\lambda_3^u\lambda_7 + (-8\bar g_3^2 +6\bar{Y}_t^2 + 2\bar{Y}_{t_R}^2 +2\bar{Y}_{b_R}^2 )\lambda^u_1 -4\bar{Y}_{t_R}^2 \bar{Y}_t^2 \,,  \label{eq:lambdau1}
\\
16\pi^2\frac{d}{dt} \lambda_2^u &=& 8\lambda_1^u \lambda_2^u +4(\lambda_2^u)^2 +4\lambda_2^u\lambda_4 +12\lambda_2^u\lambda_5 
\nonumber \\
&& + (-8\bar g_3^2 +6\bar{Y}_t^2 + 2\bar{Y}_{t_R}^2 +2\bar{Y}_{b_R}^2 )\lambda_2^u \,, \label{eq:lambdau2}
\\
16\pi^2\frac{d}{dt} \lambda_3^u &=& 
12\lambda_1^u \lambda_6 + 6 \lambda_2^u \lambda_6 
+4 \lambda_1^u \lambda_7 +2\lambda_2^u\lambda_7 +4 (\lambda_3^u)^2 
+16\lambda_3^u \lambda_8 
\nonumber \\
&& + (-8\bar g_3^2 +6\bar{Y}_t^2 + 4\bar{Y}_{q_{3L}}^2 )\lambda_3^u -4\bar{Y}_{q_{3L}}^2\bar{Y}_t^2 \,,  
\\
16\pi^2\frac{d}{dt} \lambda_1^d &=& 
4(\lambda_1^d)^2 +2(\lambda_2^d)^2
+28\lambda_1^d\lambda_4 +20\lambda_1^d\lambda_5 +12\lambda_2^d\lambda_4 + 4\lambda_2^d\lambda_5
\nonumber \\ 
&&  +6\lambda_3^d\lambda_6 
+2\lambda_3^d\lambda_7 + (-8\bar g_3^2 +6\bar{Y}_b^2 + 2\bar{Y}_{t_R}^2 +2\bar{Y}_{b_R}^2 )\lambda_1^d-4\bar{Y}_{b_R}^2\bar{Y}_b^2 \,, \label{eq:lambdad1}
\\
16\pi^2\frac{d}{dt} \lambda_2^d &=& 
8\lambda_1^d \lambda_2^d +4(\lambda_2^d)^2 
+4\lambda_2^d\lambda_4 +12\lambda_2^d\lambda_5 
\nonumber \\
&& + (-8\bar g_3^2 +6\bar{Y}_b^2 + 2\bar{Y}_{t_R}^2 +2\bar{Y}_{b_R}^2 )
\lambda_2^d \,, \label{eq:lambdad2}
\\
16\pi^2\frac{d}{dt} \lambda_3^d &=& 
12\lambda_1^d \lambda_6 + 6 \lambda_2^d \lambda_6 
+4 \lambda_1^d \lambda_7 +2\lambda_2^d\lambda_7 +4 (\lambda_3^d)^2 
+16\lambda_3^d \lambda_8 
\nonumber \\
&& + (-8\bar g_3^2 +6\bar{Y}_b^2 + 4\bar{Y}_{q_{3L}}^2 )\lambda^d_3 -4\bar{Y}_{q_{3L}}^2\bar{Y}_b^2\,, 
\\\nonumber
16\pi^2\frac{d}{dt} \lambda_4 &=& 
2 (\lambda_1^u)^2 +2\lambda_1^u \lambda_2^u +2 (\lambda_1^d)^2 +2\lambda_1^d \lambda_2^d +40\lambda_4^2 +40\lambda_4 \lambda_5 + 12\lambda_5^2 + 3\lambda_6^2 
\nonumber \\ 
&& +2\lambda_6 \lambda_7 +(-16 \bar g_3^2 +4(\bar{Y}_{t_R}^2 + \bar{Y}_{b_R}^2 ) ) \lambda_4  +\frac{11}{12}\bar g_3^4\,,\label{eq:lambda4}
\\ 
16\pi^2\frac{d}{dt} \lambda_5 &=& 
(\lambda_2^u)^2 + (\lambda_2^d)^2 +24\lambda_4 \lambda_5 + 20\lambda_5^2 + \lambda_7^2 +(-16 \bar g_3^2 +4(\bar{Y}_{t_R}^2 + \bar{Y}_{b_R}^2 ) ) \lambda_5 
\nonumber \\ 
&& -2(\bar{Y}_{t_R}^4+\bar{Y}_{b_R}^4) +\frac{5}{4} \bar g_3^4\,,\label{eq:lambda5}
\\ 
16\pi^2\frac{d}{dt} \lambda_6 &=& 
(4\lambda_1^u+2\lambda_2^u)\lambda_3^u +(4\lambda_1^d+2\lambda_2^d)\lambda_3^d+ 28\lambda_4 \lambda_6 
+ 8\lambda_4 \lambda_7 +20\lambda_5\lambda_6 + 4\lambda_5\lambda_7 
\nonumber \\
&& +4\lambda_6^2 + 2\lambda_7^2 +16 \lambda_6 \lambda_8 +4\lambda_7\lambda_8+ (-16 \bar g_3^2 +2\bar{Y}_{t_R}^2 +2\bar{Y}_{b_R}^2 + 4\bar{Y}_{q_{3L}}^2 ) \lambda_6 
\nonumber \\
&& -4\bar{Y}_{t_R}^2 \bar{Y}_{q_{3L}}^2 +\frac{11}{6}\bar g_3^4\,, \label{eq:lambda6}
\\
16\pi^2\frac{d}{dt} \lambda_7 &=& 
4\lambda_4 \lambda_7 +8\lambda_5\lambda_7 + 8\lambda_6\lambda_7 
+6\lambda_7^2 +4\lambda_7\lambda_8
\nonumber \\
&& + (-16 \bar g_3^2 +2\bar{Y}_{t_R}^2 +2\bar{Y}_{b_R}^2 + 4\bar{Y}_{q_{3L}}^2 ) \lambda_7 +\frac{5}{2} \bar g_3^4\,, \label{eq:lambda7}
\\
16\pi^2\frac{d}{dt} \lambda_8 &=& 
2 (\lambda_3^u)^2+2 (\lambda_3^d)^2 +6\lambda_6^2 +4\lambda_6 \lambda_7
+2\lambda_7^2 +28\lambda_8^2 +(-16 \bar g_3^2 +8\bar{Y}_{q_{3L}}^2 ) \lambda_8
\nonumber \\ 
&&  -4\bar{Y}_{q_{3L}}^4 +\frac{13}{6} \bar g_3^4\,. 
\end{eqnarray}
Note that in all equations above we assumed real parameters. However, all formula can be easily generalized to the complex case by simply replacing a square by the absolute value squared.

By integrating these RGEs from $M$ to the stop mass scale $m_{\tilde{t}}$, we obtain the $O(\alpha_3,Y_{t,b})$ contributions enhanced by $\log(M/m_{\tilde{t}})$. 

\subsection{Stop masses}
In the effective theory, the stop mass matrix in the ($\tilde{t}_L$, $\tilde{t}_R$) basis reads
\begin{equation}
\begin{aligned}
\renewcommand{\arraystretch}{2.0}
	\bar{\cal M}_{\tilde{t}}^2 &= 
	\left(\begin{array}{cc}		
		\bar m_{\tilde Q}^2 + v_u^2  \lambda_1^u+v_d^2(\lambda_1^d+\lambda_2^d) & v_u \bar A_t^* -  v_d \bar \mu_t^*  \\
 v_u \bar A_t -  v_d \bar \mu_t  &\bar m_{\tilde t}^2 + v_u^2 \lambda_3^u+v_d^2 \lambda_3^d
	\end{array}	\right)\,, \\
	\label{up-squark-mass-matrix,eff}
\end{aligned}
\end{equation}
where $v_{u,d}=\langle H^0_{u,d}\rangle$ are the vacuum expectation values of the Higgs scalars. By diagonalizing this matrix one obtains the stop masses and the stop mixing angle, both in the $\overline{\rm MS}$ scheme. 
These masses are closely related to the left-handed sbottom mass 
\begin{equation}
M_{\tilde{b}_L}^2 = \bar{m}_{\tilde Q}^2+ v_u^2 (\lambda_1^u +\lambda_2^u) + v_d^2 \lambda_1^d, 
\end{equation}
by SU(2) gauge symmetry. 

\begin{figure}[t]
\centering
\includegraphics[width=0.9\textwidth]{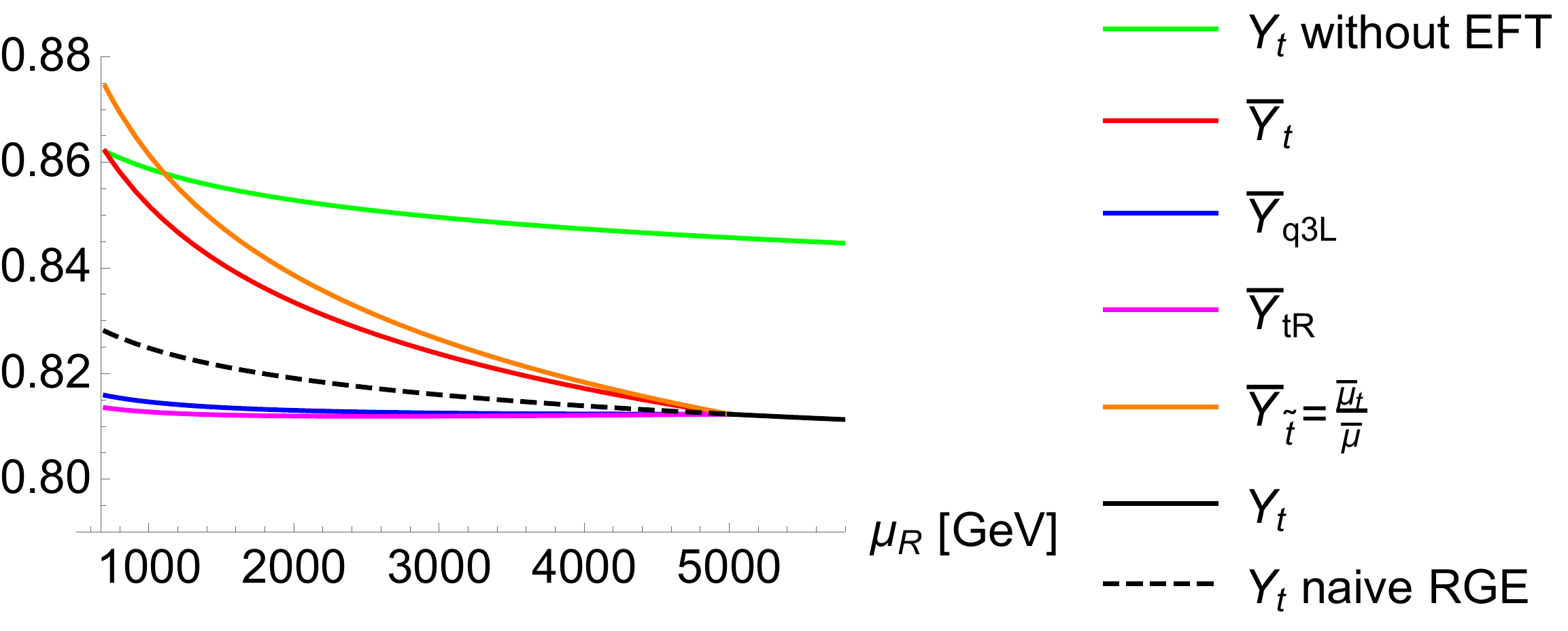}
\caption{Evolution of the Yukawa coupling $Y_t$ in the naive approach without using an EFT (green) compared to the various Higgs/Higgsino-stop/top couplings in the EFT for $M=5\,$TeV and $\tan\beta=50$ as a function of the renormalization scale ${\mu_R}$. Note that the only numerically sizable impact of $\tan\beta=50$ is the splitting between the $\bar{Y}_{t_R}$ and $\bar{Y}_{q_{3L}}$. The initial condition of the Yukawa coupling is determined by the requirement that $v_u Y_t=m_t=150$~GeV at the stop scale which we choose here to be $500\,$GeV. $\bar Y_{\tilde t}=\bar\mu_t/\bar \mu$ shows the evolution of the $\tilde{t}-\tilde{t}-H_d$ coupling relative to the Higgsino mass term $\bar{\mu}$ in the EFT. We also show the projected evolution of $Y_t$ below the scale $M$ (black-dashed) in the MSSM  RGE for the boundary condition $Y_t(M)=\bar{Y}_t(M)$. Note that above the scale $M$ SUSY is restored, so that there is only one Yukawa coupling $Y_t$ (black).}
\label{fig:Yt}
\end{figure}

\section{Numerical Analysis}\label{sec:numerics}

From the previous analysis, we can see that, by integrating out the gluino and the squarks of the first two generations, parameters which were originally related via SUSY in the full MSSM, do not evolve anymore in the same way in the EFT. Let us illustrate this effect with two examples where striking differences between the EFT approach and the full MSSM emerge. Here we set the input parameters as $M=5$~TeV, the stop mass scale $m_{\tilde{t}}=700$~GeV, running top mass $m_t(m_{\tilde{t}})=\bar{Y}_t(m_{\tilde{t}})v_u=150$~GeV, $\alpha_3(m_{\tilde{t}})=0.1$, and $\tan\beta=v_u/v_d=50$. Furthermore, we have chosen the massive parameters such that the collider constraints for the Higgs mass and the stop and sbottom masses are fulfilled. This can be achieved by using the values: $\bar m_{\tilde t}(m_{\tilde t})=800$~GeV, $\bar m_{\tilde Q}(m_{\tilde t})=900$~GeV, $\bar A_t(m_{\tilde t})=1200$~GeV which lead to a one-loop mass of $125$~GeV for the lightest Higgs, a light stop of $700$~GeV and a sbottom mass of about $900$~GeV.

\begin{figure}[t]
\centering
\includegraphics[width=0.9\textwidth]{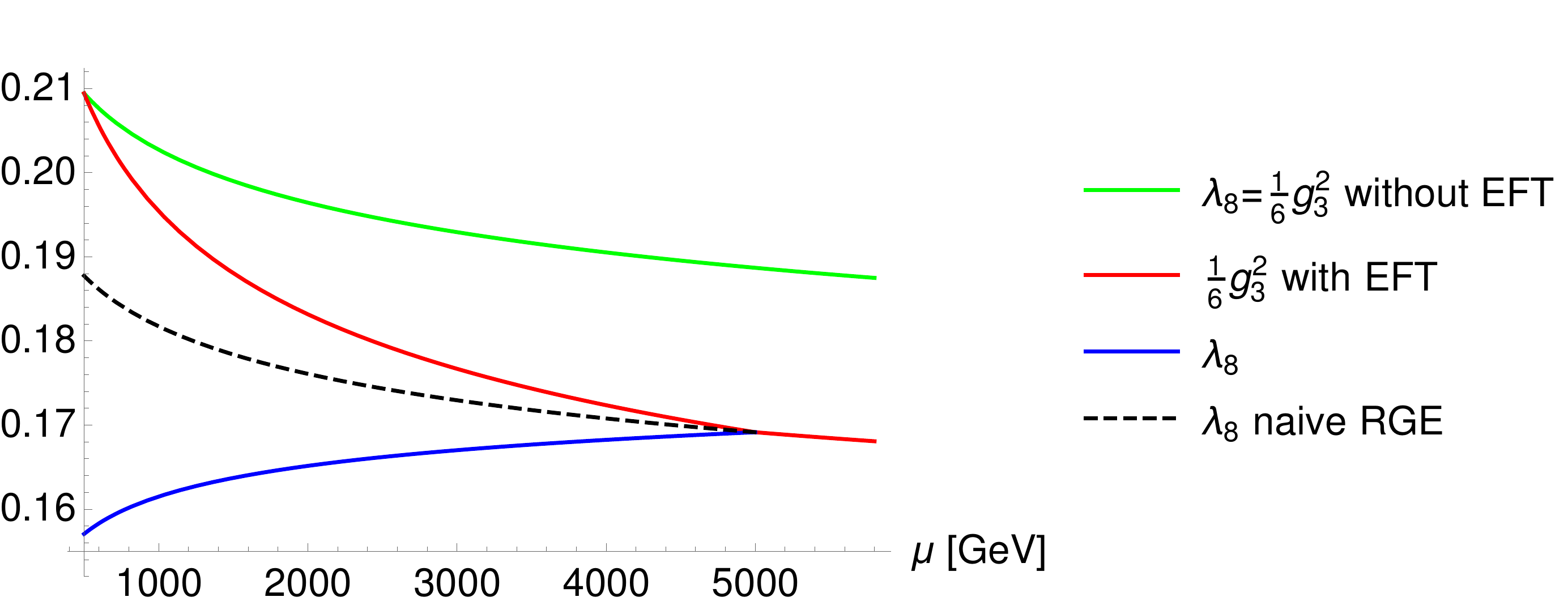}
\caption{Evolution of the quartic coupling to right-handed stops in the naive approach with the MSSM RGE (green)  compared to the EFT approach, where $\alpha_3=0.1$ at the stop scale. Note that the SUSY relation $\lambda_8=\frac{1}{6}g_3^2$ holds only at the scale $M$ in the EFT. 
The dotted-black line shows the projected evolution of $\lambda_8$ for the boundary condition $\lambda_8(M)=\frac{1}{6}\bar{g}_3^2(M)$ with the naive RGE of the full MSSM. Note that above the scale $M$ SUSY is restored, $\lambda_8=1/6g_3^2$ and evolves like $g_3^2$ in the full MSSM.}
\label{fig:lambda8}
\end{figure}

\begin{itemize}
	\item The top Yukawa coupling $Y_t$\newline
In the full MSSM, the Yukawa coupling $Y_t$ of the superpotential enters top-top-Higgs, stop-stop-Higgs couplings as well as stop-squark-Higgsino couplings in the same way. However, in the EFT these couplings are independent quantities and they evolve differently below the scale $M$. This is depicted in Fig.~\ref{fig:Yt}, where the evolution of $Y_t$ in the naive approach using MSSM RGE is compared to those of $\bar Y_t$, $\bar Y_{ q_{3L}}$, $\bar Y_{t_R}$ and $\bar Y_{\tilde t}\equiv\bar\mu_t/\bar \mu$ in the EFT. When the values of $\bar Y_t$ and $Y_t$ are determined at the stop mass scale to give the SM running top mass, their values at the scale $M$ are quite different. Note that these couplings are dimensionless and therefore do not depend on the choice of the parameters for $\bar m_{\tilde t},\,\bar m_{\tilde Q}$ and $\bar A_t$.
	\item The quartic coupling of right-handed stops $\lambda_8$\newline
In the full MSSM the quartic coupling of right-handed stops $\lambda_8$ is given by $\frac{1}{6}g_3^2$ by SUSY relation and evidently also evolves in the same way as $\frac{1}{6}g_3^2$. However, in the EFT $\lambda_8$ and $\bar{g}_3^2$ follow different RGEs below the scale $M$, as seen in Fig.~\ref{fig:lambda8}. 
The relative difference at the scale $m_{\tilde t}$ amounts to roughly 30\%. Again, since $\lambda_8$ has no mass dimension, its running does not depend on $\bar m_{\tilde t},\,\bar m_{\tilde Q}$ and $\bar A_t$.
\end{itemize}

Among the quartic scalar couplings $\lambda_{1-8}$, the running of $\lambda_8$ in the EFT exhibits the largest deviation from the one in the full theory. This is due to symmetry factors, leading to large coefficients of the box diagrams and self-couplings which are responsible for a change in sign on the $g_3^4$-dependence. The deviations of the other couplings $\lambda_{1-7}$ from the ones in the full theory are either positive or negative, but are smaller than 20\% for our parameter set. 
We therefore do not show the figures of their runnings here.

\section{Conclusions}\label{sec:conclusions}
In this article, we constructed an effective theory of the stop sector obtained from the full MSSM by integrating out the first and second generation of squarks and the gluino (which we assume to have a common mass of the order $M$). We computed the matching effects for the dimensionful quantities which are enhanced by powers of $M$ at $O(\alpha_3,Y_{t,b}^2)$. In addition, we obtained the complete $O(\alpha_3,Y_{t,b}^2)$ RGEs of the couplings within the EFT. In the numerical analysis we highlighted that couplings which are related via SUSY identities within the full MSSM have different RGEs within the EFT, which can lead to sizable differences. We illustrated this effect for the top Yukawa couplings and the quartic coupling of right-handed stops, finding differences up to 30\% between the EFT and the naive approach. Such deviations could play a role in future test of the stop-stop or stop-Higgs interactions which also enter the calculation of the Higgs mass.

\begin{acknowledgments}
J.A. and C.G. acknowledge the support from the Swiss National Science Foundation. The work of A.C. was supported by a Marie Curie Intra-European Fellowship of the European Community's 7th Framework Programme under contract number (PIEF-GA-2012-326948) and by an Ambizione Grant of the Swiss National Science Foundation.
\end{acknowledgments}

\vspace{1cm}
\appendix

{\bf Appendix}
\subsection{RGEs of the full MSSM}
Here we recall the RGEs of the parameters in the full MSSM, again taking into account $O(\alpha_3)$ and $O(Y_{t,b}^2)$ effects.
\begin{eqnarray}
16\pi^2\frac{d}{dt} g_3 &=& -3 g_3^3\,,\\
16\pi^2\frac{d}{dt}Y_t &=& Y_t \left[ -\frac{16}{3} g_3^2 +6Y_t^2 +Y_b^2 \right] \,, \\
16\pi^2\frac{d}{dt}Y_b &=& Y_b \left[-\frac{16}{3} g_3^2 +Y_t^2 + 6Y_b^2 \right] \,,
\end{eqnarray}
\begin{eqnarray}
16\pi^2\frac{d}{dt}m_{H_u}^2 &=&  6\left(Y_t^2(m_{H_u}^2 + m_{\tilde{Q}}^2 +  m_{\st}^2) + A_t^2\right)\,,\\
16\pi^2\frac{d}{dt} m_{H_d}^2 &=& 6 (Y_b^2(m_{H_d}^2+ m_{\tilde{Q}}^2 +m_{\tilde b_R}^2)+A_b^2) \,,\\ 
16\pi^2\frac{d}{dt} m_{H_dH_u}^2 &=&  3(Y_t^2+  Y_b^2 ) m_{H_dH_u}^2+6(Y_t A_t+Y_b A_b)\mu \,,\\
16\pi^2 \frac{d}{dt} m_{\tilde{Q}}^2 &=& -\frac{32}{3}g_3^2m_{\tilde g}^2 +2Y_t^2(m_{\tilde{Q}}^2+m_{H_u}^2+m_{\st}^2)\\ \nonumber && +2Y_b^2(m_{\tilde{Q}}^2+m_{H_d}^2+m_{\tilde b_R}^2)+2(A_t^2+A_b^2)\, ,\\
16\pi^2 \frac{d}{dt}  m_{\st}^2 &=& -\frac{32}{3}g_3^2m_{\tilde g}^2 +4Y_t^2(m_{\tilde{Q}}^2+m_{\tilde{t}}^2+m_{H_u}^2)+4A_t^2\, ,\\
16\pi^2 \frac{d}{dt}  m_{\tilde b_R}^2 &=& -\frac{32}{3}g_3^2m_{\tilde g}^2 +4Y_b^2(m_{\tilde{Q}}^2+m_{\tilde{b}}^2+m_{H_d}^2)+4A_b^2\, ,\\
16\pi^2 \frac{d}{dt}  \mu &=&  3(Y_t^2+Y_b^2) \mu \,,\\ 
16\pi^2\frac{d}{dt}A_t &=& A_t \left[ -\frac{16}{3} g_3^2 +18Y_t^2+Y_b^2 \right]+2Y_t Y_bA_b+\frac{32}{3}g_3^2m_{\tilde g}Y_t\, ,\\
16\pi^2\frac{d}{dt}A_b &=& A_b \left[ -\frac{16}{3} g_3^2 +Y_t^2+18Y_b^2 \right]+2Y_t Y_bA_t+\frac{32}{3}g_3^2m_{\tilde g}Y_b\, ,\\
16\pi^2\frac{d}{dt}m_{\tilde g} &=& -6g_3^2m_{\tilde g}\, .
\end{eqnarray}


\newpage
\begin{thebibliography}{99}
\bibliographystyle{hieeetr}
\bibitem{Dimopoulos:1995mi} 
  S.~Dimopoulos and G.~F.~Giudice,
  Phys.\ Lett.\ B {\bf 357}, 573 (1995)
  doi:10.1016/0370-2693(95)00961-J
  [hep-ph/9507282].



\bibitem{Giudice:2004tc} 
  G.~F.~Giudice and A.~Romanino,
  Nucl.\ Phys.\ B {\bf 699}, 65 (2004)
  Erratum: [Nucl.\ Phys.\ B {\bf 706}, 487 (2005)]
  doi:10.1016/j.nuclphysb.2004.11.048, 10.1016/j.nuclphysb.2004.08.001
  [hep-ph/0406088].



\bibitem{Inoue:1982pi} 
  K.~Inoue, A.~Kakuto, H.~Komatsu and S.~Takeshita,
  Prog.\ Theor.\ Phys.\  {\bf 68}, 927 (1982)
  Erratum: [Prog.\ Theor.\ Phys.\  {\bf 70}, 330 (1983)].
  doi:10.1143/PTP.68.927



\bibitem{Inoue:1983pp} 
  K.~Inoue, A.~Kakuto, H.~Komatsu and S.~Takeshita,
  Prog.\ Theor.\ Phys.\  {\bf 71}, 413 (1984).
  doi:10.1143/PTP.71.413



\bibitem{Derendinger:1983bz} 
  J.~P.~Derendinger and C.~A.~Savoy,
  Nucl.\ Phys.\ B {\bf 237}, 307 (1984).
  doi:10.1016/0550-3213(84)90162-7



\bibitem{Gato:1984ya} 
  B.~Gato, J.~Leon, J.~Perez-Mercader and M.~Quiros,
  Nucl.\ Phys.\ B {\bf 253}, 285 (1985).
  doi:10.1016/0550-3213(85)90532-2



\bibitem{Falck:1985aa} 
  N.~K.~Falck,
  Z.\ Phys.\ C {\bf 30}, 247 (1986).
  doi:10.1007/BF01575432



\bibitem{Martin:1993zk} 
  S.~P.~Martin and M.~T.~Vaughn,
  Phys.\ Rev.\ D {\bf 50}, 2282 (1994)
  Erratum: [Phys.\ Rev.\ D {\bf 78}, 039903 (2008)]
  doi:10.1103/PhysRevD.50.2282, 10.1103/PhysRevD.78.039903
  [hep-ph/9311340].



\bibitem{Aad:2012tfa} 
  G.~Aad {\it et al.} [ATLAS Collaboration],
  Phys.\ Lett.\ B {\bf 716}, 1 (2012)
  doi:10.1016/j.physletb.2012.08.020
  [arXiv:1207.7214 [hep-ex]].



\bibitem{Chatrchyan:2012ufa} 
  S.~Chatrchyan {\it et al.} [CMS Collaboration],
  Phys.\ Lett.\ B {\bf 716}, 30 (2012)
  doi:10.1016/j.physletb.2012.08.021
  [arXiv:1207.7235 [hep-ex]].



\bibitem{Ellis:1990nz} 
  J.~R.~Ellis, G.~Ridolfi and F.~Zwirner,
  Phys.\ Lett.\ B {\bf 257}, 83 (1991).
  doi:10.1016/0370-2693(91)90863-L



\bibitem{Okada:1990vk} 
  Y.~Okada, M.~Yamaguchi and T.~Yanagida,
  Prog.\ Theor.\ Phys.\  {\bf 85}, 1 (1991).
  doi:10.1143/PTP.85.1



\bibitem{Haber:1990aw} 
  H.~E.~Haber and R.~Hempfling,
  Phys.\ Rev.\ Lett.\  {\bf 66}, 1815 (1991).
  doi:10.1103/PhysRevLett.66.1815



\bibitem{Ellwanger:2009dp} 
  U.~Ellwanger, C.~Hugonie and A.~M.~Teixeira,
  Phys.\ Rept.\  {\bf 496}, 1 (2010)
  doi:10.1016/j.physrep.2010.07.001
  [arXiv:0910.1785 [hep-ph]].



\bibitem{Hall:2011aa} 
  L.~J.~Hall, D.~Pinner and J.~T.~Ruderman,
  JHEP {\bf 1204}, 131 (2012)
  doi:10.1007/JHEP04(2012)131
  [arXiv:1112.2703 [hep-ph]].



\bibitem{Chala:2017jgg} 
  M.~Chala, A.~Delgado, G.~Nardini and M.~Quiros,
  JHEP {\bf 1704}, 097 (2017)
  doi:10.1007/JHEP04(2017)097
  [arXiv:1702.07359 [hep-ph]].



\bibitem{Batra:2003nj} 
  P.~Batra, A.~Delgado, D.~E.~Kaplan and T.~M.~P.~Tait,
  JHEP {\bf 0402}, 043 (2004)
  doi:10.1088/1126-6708/2004/02/043
  [hep-ph/0309149].



\bibitem{Espinosa:1998re} 
  J.~R.~Espinosa and M.~Quiros,
  Phys.\ Rev.\ Lett.\  {\bf 81}, 516 (1998)
  doi:10.1103/PhysRevLett.81.516
  [hep-ph/9804235].



\bibitem{Djouadi:2005gj} 
  A.~Djouadi,
  Phys.\ Rept.\  {\bf 459}, 1 (2008)
  doi:10.1016/j.physrep.2007.10.005
  [hep-ph/0503173].



\bibitem{Brummer:2012ns} 
  F.~Brummer, S.~Kraml and S.~Kulkarni,
  JHEP {\bf 1208}, 089 (2012)
  doi:10.1007/JHEP08(2012)089
  [arXiv:1204.5977 [hep-ph]].



\bibitem{Wymant:2012zp} 
  C.~Wymant,
  Phys.\ Rev.\ D {\bf 86}, 115023 (2012)
  doi:10.1103/PhysRevD.86.115023
  [arXiv:1208.1737 [hep-ph]].



\bibitem{Aad:2014qaa} 
  G.~Aad {\it et al.} [ATLAS Collaboration],
  JHEP {\bf 1406}, 124 (2014)
  doi:10.1007/JHEP06(2014)124
  [arXiv:1403.4853 [hep-ex]].


\bibitem{ATLAS-CONF-2016-050} 
  The ATLAS collaboration [ATLAS Collaboration],
  ATLAS-CONF-2016-050.



\bibitem{Chatrchyan:2013xna} 
  S.~Chatrchyan {\it et al.} [CMS Collaboration],
  Eur.\ Phys.\ J.\ C {\bf 73}, no. 12, 2677 (2013)
  doi:10.1140/epjc/s10052-013-2677-2
  [arXiv:1308.1586 [hep-ex]].



\bibitem{Cheng:2016mcw} 
  H.~C.~Cheng, C.~Gao, L.~Li and N.~A.~Neill,
  JHEP {\bf 1605}, 036 (2016)
  doi:10.1007/JHEP05(2016)036
  [arXiv:1604.00007 [hep-ph]].



\bibitem{Porod:1996at} 
  W.~Porod and T.~Wohrmann,
  Phys.\ Rev.\ D {\bf 55}, 2907 (1997)
  Erratum: [Phys.\ Rev.\ D {\bf 67}, 059902 (2003)]
  doi:10.1103/PhysRevD.67.059902, 10.1103/PhysRevD.55.2907
  [hep-ph/9608472].



\bibitem{ATLAS-CONF-2016-076} 
  The ATLAS collaboration [ATLAS Collaboration],
  ATLAS-CONF-2016-076.



\bibitem{Aaboud:2016tnv} 
  M.~Aaboud {\it et al.} [ATLAS Collaboration],
  Phys.\ Rev.\ D {\bf 94}, no. 3, 032005 (2016)
  doi:10.1103/PhysRevD.94.032005
  [arXiv:1604.07773 [hep-ex]].



\bibitem{Aebischer:2014lfa} 
  J.~Aebischer, A.~Crivellin and C.~Greub,
  Phys.\ Rev.\ D {\bf 91}, no. 3, 035010 (2015)
  doi:10.1103/PhysRevD.91.035010
  [arXiv:1410.8459 [hep-ph]].



\bibitem{Grober:2014aha} 
  R.~Gröber, M.~M.~Mühlleitner, E.~Popenda and A.~Wlotzka,
  Eur.\ Phys.\ J.\ C {\bf 75}, 420 (2015)
  doi:10.1140/epjc/s10052-015-3626-z
  [arXiv:1408.4662 [hep-ph]].



\bibitem{Aad:2014nra} 
  G.~Aad {\it et al.} [ATLAS Collaboration],
  Phys.\ Rev.\ D {\bf 90}, no. 5, 052008 (2014)
  doi:10.1103/PhysRevD.90.052008
  [arXiv:1407.0608 [hep-ex]].



\bibitem{Macaluso:2015wja} 
  S.~Macaluso, M.~Park, D.~Shih and B.~Tweedie,
  JHEP {\bf 1603}, 151 (2016)
  doi:10.1007/JHEP03(2016)151
  [arXiv:1506.07885 [hep-ph]].



\bibitem{Belyaev:2015gna} 
  A.~Belyaev, V.~Sanz and M.~Thomas,
  JHEP {\bf 1601}, 102 (2016)
  doi:10.1007/JHEP01(2016)102
  [arXiv:1510.07688 [hep-ph]].



\bibitem{Kobakhidze:2015scd} 
  A.~Kobakhidze, N.~Liu, L.~Wu, J.~M.~Yang and M.~Zhang,
  Phys.\ Lett.\ B {\bf 755}, 76 (2016)
  doi:10.1016/j.physletb.2016.02.003
  [arXiv:1511.02371 [hep-ph]].



\bibitem{Crivellin:2016rdu} 
  A.~Crivellin, U.~Haisch and L.~C.~Tunstall,
  JHEP {\bf 1609}, 080 (2016)
  doi:10.1007/JHEP09(2016)080
  [arXiv:1604.00440 [hep-ph]].



\bibitem{ATLAS:2016lsr} 
  The ATLAS collaboration [ATLAS Collaboration],
  ATLAS-CONF-2016-054.



\bibitem{Sakuma:2016nxo} 
  T.~Sakuma [CMS Collaboration],
  PoS LHCP {\bf 2016}, 145 (2017)
  [arXiv:1609.07445 [hep-ex]].



\bibitem{ATLAS:2017qih} 
  The ATLAS collaboration [ATLAS Collaboration],
  ATLAS-CONF-2017-038.



\bibitem{Sirunyan:2017kiw} 
  A.~M.~Sirunyan {\it et al.} [CMS Collaboration],
  arXiv:1707.07274 [hep-ex].



\bibitem{Aad:2015jqa} 
  G.~Aad {\it et al.} [ATLAS Collaboration],
  Eur.\ Phys.\ J.\ C {\bf 75}, no. 5, 208 (2015)
  doi:10.1140/epjc/s10052-015-3408-7
  [arXiv:1501.07110 [hep-ex]].



\bibitem{ATLAS:2016uwq} 
  The ATLAS collaboration [ATLAS Collaboration],
  ATLAS-CONF-2016-096.



\bibitem{Arina:2016rbb} 
  C.~Arina, M.~Chala, V.~Martin-Lozano and G.~Nardini,
  JHEP {\bf 1612}, 149 (2016)
  doi:10.1007/JHEP12(2016)149
  [arXiv:1610.03822 [hep-ph]].



\bibitem{Khachatryan:2014mma} 
  V.~Khachatryan {\it et al.} [CMS Collaboration],
  Phys.\ Rev.\ D {\bf 90}, no. 9, 092007 (2014)
  doi:10.1103/PhysRevD.90.092007
  [arXiv:1409.3168 [hep-ex]].



\bibitem{CMS:2016gvu} 
  CMS Collaboration [CMS Collaboration],
  CMS-PAS-SUS-16-024.



\bibitem{Aad:2014vma} 
  G.~Aad {\it et al.} [ATLAS Collaboration],
  JHEP {\bf 1405}, 071 (2014)
  doi:10.1007/JHEP05(2014)071
  [arXiv:1403.5294 [hep-ex]].



\bibitem{Bharucha:2013epa} 
  A.~Bharucha, S.~Heinemeyer and F.~von der Pahlen,
  Eur.\ Phys.\ J.\ C {\bf 73}, no. 11, 2629 (2013)
  doi:10.1140/epjc/s10052-013-2629-x
  [arXiv:1307.4237 [hep-ph]].



\bibitem{Martin:2014qra} 
  T.~A.~W.~Martin and D.~Morrissey,
  JHEP {\bf 1412}, 168 (2014)
  doi:10.1007/JHEP12(2014)168
  [arXiv:1409.6322 [hep-ph]].



\bibitem{Calibbi:2014lga} 
  L.~Calibbi, J.~M.~Lindert, T.~Ota and Y.~Takanishi,
  JHEP {\bf 1411}, 106 (2014)
  doi:10.1007/JHEP11(2014)106
  [arXiv:1410.5730 [hep-ph]].



\bibitem{Chakraborti:2015mra} 
  M.~Chakraborti, U.~Chattopadhyay, A.~Choudhury, A.~Datta and S.~Poddar,
  JHEP {\bf 1511}, 050 (2015)
  doi:10.1007/JHEP11(2015)050
  [arXiv:1507.01395 [hep-ph]].



\bibitem{ATLAS:2014kua} 
  The ATLAS collaboration [ATLAS Collaboration],
  ATLAS-CONF-2014-010.



\bibitem{Khachatryan:2014jya} 
  V.~Khachatryan {\it et al.} [CMS Collaboration],
  Phys.\ Rev.\ D {\bf 90}, 112013 (2014)
  doi:10.1103/PhysRevD.90.112013
  [arXiv:1410.2751 [hep-ex]].



\bibitem{Celis:2013rcs} 
  A.~Celis, V.~Ilisie and A.~Pich,
  JHEP {\bf 1307}, 053 (2013)
  doi:10.1007/JHEP07(2013)053
  [arXiv:1302.4022 [hep-ph]].



\bibitem{Chiang:2013ixa} 
  C.~W.~Chiang and K.~Yagyu,
  JHEP {\bf 1307}, 160 (2013)
  doi:10.1007/JHEP07(2013)160
  [arXiv:1303.0168 [hep-ph]].



\bibitem{Chen:2013rba} 
  C.~Y.~Chen, S.~Dawson and M.~Sher,
  Phys.\ Rev.\ D {\bf 88}, 015018 (2013)
  Erratum: [Phys.\ Rev.\ D {\bf 88}, 039901 (2013)]
  doi:10.1103/PhysRevD.88.015018, 10.1103/PhysRevD.88.039901
  [arXiv:1305.1624 [hep-ph]].



\bibitem{Craig:2013hca} 
  N.~Craig, J.~Galloway and S.~Thomas,
  arXiv:1305.2424 [hep-ph].



\bibitem{Wang:2014lta} 
  L.~Wang and X.~F.~Han,
  JHEP {\bf 1411}, 085 (2014)
  doi:10.1007/JHEP11(2014)085
  [arXiv:1404.7437 [hep-ph]].



\bibitem{Grinstein:2013npa} 
  B.~Grinstein and P.~Uttayarat,
  JHEP {\bf 1306}, 094 (2013)
  Erratum: [JHEP {\bf 1309}, 110 (2013)]
  doi:10.1007/JHEP09(2013)110, 10.1007/JHEP06(2013)094
  [arXiv:1304.0028 [hep-ph]].



\bibitem{Han:2017pfo} 
  L.~Wang, F.~Zhang and X.~F.~Han,
  Phys.\ Rev.\ D {\bf 95}, no. 11, 115014 (2017)
  doi:10.1103/PhysRevD.95.115014
  [arXiv:1701.02678 [hep-ph]].



\bibitem{Aaboud:2016cre} 
  M.~Aaboud {\it et al.} [ATLAS Collaboration],
  Eur.\ Phys.\ J.\ C {\bf 76}, no. 11, 585 (2016)
  doi:10.1140/epjc/s10052-016-4400-6
  [arXiv:1608.00890 [hep-ex]].



\bibitem{Khachatryan:2014wca} 
  V.~Khachatryan {\it et al.} [CMS Collaboration],
  JHEP {\bf 1410}, 160 (2014)
  doi:10.1007/JHEP10(2014)160
  [arXiv:1408.3316 [hep-ex]].



\bibitem{ArkaniHamed:2004yi} 
  N.~Arkani-Hamed, S.~Dimopoulos, G.~F.~Giudice and A.~Romanino,
  Nucl.\ Phys.\ B {\bf 709}, 3 (2005)
  doi:10.1016/j.nuclphysb.2004.12.026
  [hep-ph/0409232].



\bibitem{Muhlleitner:2008yw} 
  M.~Muhlleitner, H.~Rzehak and M.~Spira,
  JHEP {\bf 0904}, 023 (2009)
  doi:10.1088/1126-6708/2009/04/023
  [arXiv:0812.3815 [hep-ph]].



\bibitem{Carena:2008rt} 
  M.~Carena, G.~Nardini, M.~Quiros and C.~E.~M.~Wagner,
  JHEP {\bf 0810}, 062 (2008)
  doi:10.1088/1126-6708/2008/10/062
  [arXiv:0806.4297 [hep-ph]].



\bibitem{Giudice:2011cg} 
  G.~F.~Giudice and A.~Strumia,
  Nucl.\ Phys.\ B {\bf 858}, 63 (2012)
  doi:10.1016/j.nuclphysb.2012.01.001
  [arXiv:1108.6077 [hep-ph]].



\bibitem{Pomarol:1998sd} 
  A.~Pomarol and M.~Quiros,
  Phys.\ Lett.\ B {\bf 438}, 255 (1998)
  doi:10.1016/S0370-2693(98)00979-4
  [hep-ph/9806263].



\bibitem{Dimopoulos:2014aua} 
  S.~Dimopoulos, K.~Howe and J.~March-Russell,
  Phys.\ Rev.\ Lett.\  {\bf 113}, 111802 (2014)
  doi:10.1103/PhysRevLett.113.111802
  [arXiv:1404.7554 [hep-ph]].



\bibitem{Garcia:2015sfa} 
  I.~Garcia Garcia, K.~Howe and J.~March-Russell,
  JHEP {\bf 1512}, 005 (2015)
  doi:10.1007/JHEP12(2015)005
  [arXiv:1510.07045 [hep-ph]].



\bibitem{Delgado:2016vib} 
  A.~Delgado, M.~Garcia-Pepin, G.~Nardini and M.~Quiros,
  Phys.\ Rev.\ D {\bf 94}, no. 9, 095017 (2016)
  doi:10.1103/PhysRevD.94.095017
  [arXiv:1608.06470 [hep-ph]].



\bibitem{Chankowski:1989du} 
  P.~H.~Chankowski,
  Phys.\ Rev.\ D {\bf 41}, 2877 (1990).
  doi:10.1103/PhysRevD.41.2877



\bibitem{Hikasa:1995bw} 
  K.~i.~Hikasa and Y.~Nakamura,
  Z.\ Phys.\ C {\bf 70}, 139 (1996)
  Erratum: [Z.\ Phys.\ C {\bf 71}, 356 (1996)]
  doi:10.1007/BF02906995, 10.1007/s002880050091
  [hep-ph/9501382].



\bibitem{Nojiri:1996fp} 
  M.~M.~Nojiri, K.~Fujii and T.~Tsukamoto,
  Phys.\ Rev.\ D {\bf 54}, 6756 (1996)
  doi:10.1103/PhysRevD.54.6756
  [hep-ph/9606370].



\bibitem{Cheng:1997sq} 
  H.~C.~Cheng, J.~L.~Feng and N.~Polonsky,
  Phys.\ Rev.\ D {\bf 56}, 6875 (1997)
  doi:10.1103/PhysRevD.56.6875
  [hep-ph/9706438].



\bibitem{Cheng:1997vy} 
  H.~C.~Cheng, J.~L.~Feng and N.~Polonsky,
  Phys.\ Rev.\ D {\bf 57}, 152 (1998)
  doi:10.1103/PhysRevD.57.152
  [hep-ph/9706476].



\bibitem{Nojiri:1997ma} 
  M.~M.~Nojiri, D.~M.~Pierce and Y.~Yamada,
  Phys.\ Rev.\ D {\bf 57}, 1539 (1998)
  doi:10.1103/PhysRevD.57.1539
  [hep-ph/9707244].



\bibitem{Katz:1998br} 
  E.~Katz, L.~Randall and S.~f.~Su,
  Nucl.\ Phys.\ B {\bf 536}, 3 (1998)
  doi:10.1016/S0550-3213(98)00632-4
  [hep-ph/9801416].



\bibitem{Kiyoura:1998yt} 
  S.~Kiyoura, M.~M.~Nojiri, D.~M.~Pierce and Y.~Yamada,
  Phys.\ Rev.\ D {\bf 58}, 075002 (1998)
  doi:10.1103/PhysRevD.58.075002
  [hep-ph/9803210].



\bibitem{Hempfling:1993kv} 
  R.~Hempfling,
  Phys.\ Rev.\ D {\bf 49}, 6168 (1994).
  doi:10.1103/PhysRevD.49.6168



\bibitem{Hall:1993gn} 
  L.~J.~Hall, R.~Rattazzi and U.~Sarid,
  Phys.\ Rev.\ D {\bf 50}, 7048 (1994)
  doi:10.1103/PhysRevD.50.7048
  [hep-ph/9306309].



\bibitem{Carena:1994bv} 
  M.~Carena, M.~Olechowski, S.~Pokorski and C.~E.~M.~Wagner,
  Nucl.\ Phys.\ B {\bf 426}, 269 (1994)
  doi:10.1016/0550-3213(94)90313-1
  [hep-ph/9402253].



\bibitem{Noth:2008tw} 
  D.~Noth and M.~Spira,
  Phys.\ Rev.\ Lett.\  {\bf 101}, 181801 (2008)
  doi:10.1103/PhysRevLett.101.181801
  [arXiv:0808.0087 [hep-ph]].



\bibitem{Hofer:2009xb} 
  L.~Hofer, U.~Nierste and D.~Scherer,
  JHEP {\bf 0910}, 081 (2009)
  doi:10.1088/1126-6708/2009/10/081
  [arXiv:0907.5408 [hep-ph]].



\bibitem{Crivellin:2010er} 
  A.~Crivellin,
  Phys.\ Rev.\ D {\bf 83}, 056001 (2011)
  doi:10.1103/PhysRevD.83.056001
  [arXiv:1012.4840 [hep-ph]].



\bibitem{Crivellin:2011jt} 
  A.~Crivellin, L.~Hofer and J.~Rosiek,
  JHEP {\bf 1107}, 017 (2011)
  doi:10.1007/JHEP07(2011)017
  [arXiv:1103.4272 [hep-ph]].



\bibitem{Crivellin:2012zz} 
  A.~Crivellin and C.~Greub,
  Phys.\ Rev.\ D {\bf 87}, 015013 (2013)
  Erratum: [Phys.\ Rev.\ D {\bf 87}, 079901 (2013)]
  doi:10.1103/PhysRevD.87.015013, 10.1103/PhysRevD.87.079901
  [arXiv:1210.7453 [hep-ph]].



\bibitem{Fayet:1976cr} 
  P.~Fayet and S.~Ferrara,
  Phys.\ Rept.\  {\bf 32}, 249 (1977).
  doi:10.1016/0370-1573(77)90066-7



\bibitem{Fayet:1976et} 
  P.~Fayet,
  Phys.\ Lett.\  {\bf 64B}, 159 (1976).
  doi:10.1016/0370-2693(76)90319-1



\bibitem{Haber:1984rc} 
  H.~E.~Haber and G.~L.~Kane,
  Phys.\ Rept.\  {\bf 117}, 75 (1985).
  doi:10.1016/0370-1573(85)90051-1



\bibitem{Rosiek:1995kg} 
  J.~Rosiek,
  hep-ph/9511250.



\bibitem{West:1984dg} 
  P.~C.~West,
  Phys.\ Lett.\  {\bf 137B}, 371 (1984).
  doi:10.1016/0370-2693(84)91734-9



\bibitem{Jones:1984cx} 
  D.~R.~T.~Jones and L.~Mezincescu,
  Phys.\ Lett.\  {\bf 138B}, 293 (1984).
  doi:10.1016/0370-2693(84)91663-0



\bibitem{Martin:1993yx} 
  S.~P.~Martin and M.~T.~Vaughn,
  Phys.\ Lett.\ B {\bf 318}, 331 (1993)
  doi:10.1016/0370-2693(93)90136-6
  [hep-ph/9308222].



\bibitem{Yamada:1993ga} 
  Y.~Yamada,
  Phys.\ Rev.\ Lett.\  {\bf 72}, 25 (1994)
  doi:10.1103/PhysRevLett.72.25
  [hep-ph/9308304].






\end{thebibliography}
\end{document}